\documentstyle[onecolumn,psfig]{mn}

\input{epsf}
\begin{document}
\date{}
\title[confined binary system]
      {Statistical mechanics of confined binary system: Comparison of three
       and two dimensions}
\author[Ali Nayeri]
       {Ali ~Nayeri\thanks{email: ali@iucaa.ernet.in}\\
        Inter-University Centre for Astronomy and Astrophysics, 
            Post Bag 4, Ganeshkhind, Pune 411007, INDIA}

\maketitle
\begin{abstract}
We have used a toy model to study the behaviour of confined
binary systems in 2$D$ and compare it with previously known result in 3$D$. In the case of 2$D$, in which
canonical distribution exists only above a critical temperature,
we evaluate the exact form of partition function
for this system and 
compare the exact partition function with the mean field partition
function for the case of two-particle system. In contrast of its 3$D$ counterpart, there is no phase transition here. If this system, however,
studied in microcanonical ensemble, it shows two different phases of
kinetic energy dominated with positive specific heat and potential energy dominated with negative specific heat in presence of short distance cutoff.
In absence of short distance cutoff, surprisingly, the negative specific heat region will be replaced by region of large specific heat. This feature
is completely new and there is no such a case in 3$D$. 
\end{abstract}

\section{Introduction}
The statistical behaviour of $N$ particles interacting through Newtonian gravitational forces is very different from the statistical
behaviour  of other many body systems such as neutral gases and  plasmas.
 The central feature of gravitating system, in contrast to normal many body systems, is the {\it unshielded, long range} nature of gravitational force.
 Because of this feature one of the fundamental concepts of statistical
mechanics, the extensive nature of  energy, breaks down. This, in turn,
leads to different physical descriptions for the gravitating systems in the     microcanonical and canonical distribution.

The statistical behaviour also strongly depends on the spatial dimension.
For instance, in 3$D$, the available phase volume for the system diverges and
one is forced to use short distance cutoff. However the situation in 2$D$ is 
different. In this case there is a microcanonical description for all values
of energies, through the canonical approach exists only above some critical
temperature.

We shall study some properties of these gravitating
systems by introducing a toy-model (originally used in Padmanabhan, 1990), based on a simple Hamiltonian, describing
two particles of finite size, confined inside a box. This system shows several
important properties of more complicated systems studied for example in D. Lynden-Bell and R. M. Lyndel-Bell \shortcite{L and L}. We will
study this toy model in both the microcanonical and canonical approach.

In section 2 we will overview the properties of the toy model in 3$D$,
[which was earlier done in Padmanabhan, (1990)] for providing the background needed for comparison with the 2$D$ case, which will be studied in section 3. In section 4, we compare the nature of this system in 3$D$ and 2$D$.

An interesting feature of this 2$D$ system is that the thermodynamic functions are all calculable analytically. This contrasts with the thermodynamical behaviour of 3$D$ confined binary system. By studying
the ``{\it isothermal cylinders}'' we find that these systems are remarkably similar to
the simple toy model and that the system cannot exist at $T<T_c$ where $T_c$ is
given by $T_c=(1/2)Gm^2$. We also find that by introducing a short distance
cutoff, in contrast to 3$D$ case, the specific heat of the system would
become negative in some intermediate temperatures.
 
\section{overview of 3$D$ case}
In this section we introduce and review a toy model which was first given by Padmanabhan \shortcite{pad}. In this model, by constructing the ``statistical mechanics'' of two particles of finite size confined inside a box and interacting via 3$D$ gravity are described. This system exhibits several important properties of more complicated gravitating systems in spite of the fact that it has only two particles. In particular, this system exhibits the following  two features, which seem to be generic to all gravitating systems: (i) When studied using the microcanonical ensemble, the system shows  evidence for two different phases: a high temperature phase, dominated by kinetic energy, and a low temperature phase dominated by the potential energy and stabilised by some short distance cutoff which is of non-gravitational origin. Both these phases have positive specific heat. These two phases are connected at intermediate temperatures by a region of negative specific heat; this is precisely the range in which the kinetic and potential energies of the system  are  comparable and the system is in virial equilibrium. (ii) If the same system is studied using the canonical ensemble, the intermediate region of negative specific heat is replaced by a sharp phase transition.

Then, in order to show that this toy model exhibits several important properties of more complicated gravitating systems, we shall compare the equilibrium of a gravitating system in the mean field limit, by evaluating the partition function for this system. In the absence of any short distance cutoff to gravitational interaction, the mean field  solution  is given by an isothermal sphere. It turns out that this system is {\it remarkably similar} to our  simple toy model studied earlier.
In fact, this mean field analysis confirms all the conjectures based on the
toy model.
 
\subsection{Microcanonical description}
We will begin by studying the statistical mechanics of our toy model described by the Hamiltonian

\begin{equation}
H ({\bf P, Q ; p, r}) = {\frac{{\bf P}^2}{2M}}+{\frac{{\bf p}^2}{2\mu}}-{\frac{Gm^2}{r}}, \label{ham3}
\end{equation}

\noindent where $({\bf Q, P})$ are coordinates and momenta of the centre of mass, $({\bf r, p})$ are the relative coordinates and momenta, $M = 2m$ is the total mass, $\mu = (m/2)$ is the reduced mass and $m$ is the mass of individual particles. The range of $r$ varies in the interval $(a, R)$. This is equivalent to assuming that the particles are hard spheres of radius $(a/2)$ and that the system is confined to a spherical box of radius $R$. In microcanonical distribution corresponding to this toy model (which is the relevant ensemble to the gravitating systems ) we define the entropy $S(E)$ and temperature $T(E)$ through the relations

\begin{equation}
S(E) = \ln g(E), \hspace{1cm}
T(E) = \left(\partial S(E)/{\partial E}\right)^{-1} = \left(\partial\ln g(E)/{\partial E}\right)^{-1}. \label{ts}
\end{equation}
For our system the phase volume $g(E)$ of the constant energy surface
$H = E$ is
\begin{equation}
{\frac{g(E)}{(Gm^2)^3}} = \left\{ \begin{array}{ll}
\frac{1}{3}R^3(-E)^{-1}(1+aE/Gm^2)^3,  & \mbox{$-Gm^2/a < E < -Gm^2/R$},\vspace{5 mm}\nonumber \\
\frac{1}{3}R^3(-E)^{-1}[(1+RE/Gm^2)^3 - (1+aE/Gm^2)^3],   &  \mbox{$-Gm^2/R < E < \infty$.  \label{huge1}}               
\end{array}
\right. 
\end{equation}
This function $g(E)$ is continuous and smooth at $E = (-Gm^2/R)$. 

\begin{figure}
\centerline{\hbox{\psfig{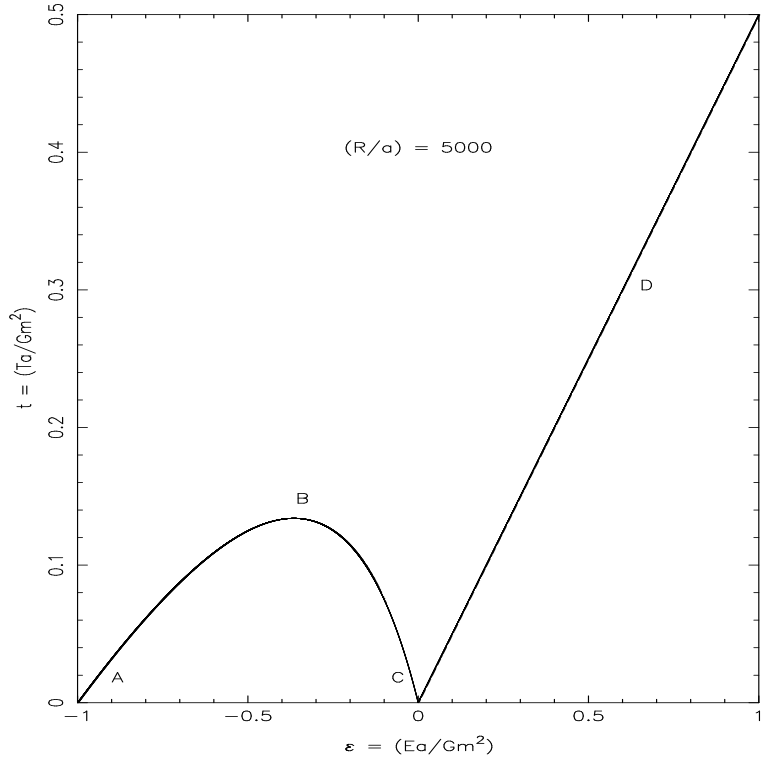}}}
\caption{ The function $T(E)$ for a confined binary system.}
\label{t1pic}
\end{figure}

\noindent Now the thermodynamic properties of the system can be analyzed from the $T(E)$ curve.

\indent In the range when $-Gm^2/a< E< -Gm^2/R$ we can write $T(E)$ in the dimensionless form 
\begin{equation}
t(\varepsilon) = \left (\frac{3}{1+\varepsilon}-\frac{1}{\varepsilon}\right )^{-1}, \label{t1}
\end{equation}
\noindent where $t = (aT/Gm^2)$ and $\varepsilon = (aE/Gm^2)$.

\indent At $\varepsilon = -1$ which corresponds to the lowest energy admissible for the system, the dimensionless temperature $t$ vanishes. It is obvious that in this case of $\varepsilon \simeq -1$, $t(\varepsilon)$ dominated by the first term of (\ref{t1}). As we increase the energy of the system, the temperature {\it increases}, which is the normal behaviour for the system. This trend continues up to
\begin{equation}
\varepsilon = {\varepsilon}_1 = -\frac{1}{2}(\sqrt{3} -1) \simeq -0.36, \label{varep1}
\end{equation}
\noindent at which point the $t(\varepsilon)$ curve reaches a maximum and turns around. As we increase the energy further the temperature {\it decreases}. {\it The system exhibits negative specific heat in this range.}

\indent As one can see from (\ref{t1}), for realistic systems, i.e., $R \gg a$
only a small region in the range of $-(Gm^2/a)$ to $-(0.36Gm^2/a)$ we will have positive specific heat; for the rest of the region the specific heat is negative. In fact the existence of the positive specific heat region is due to {\it nonzero short distance cutoff}. In the absence of this nonzero short distance cutoff, the first term in (\ref{t1}) will vanish and we will get $t \propto -\varepsilon^{-1}$ and negative specific heat in this entire region.

\indent For high energies limits, i.e., $E \geq -Gm^2/R$, the second expression in (\ref{huge1}) for $g(E)$ will give
\begin{equation}
t(\varepsilon) = \left (\frac{3\{(1+\varepsilon)^2-(R/a)[1+(R/a)\varepsilon]^2\}}{(1+\varepsilon)^3 - [1+(R/a)\varepsilon]^3}-\frac{1}{\varepsilon}\right )^{-1}. \label{t1'}
\end{equation}
\noindent This function will match with (\ref{t1}) at $\varepsilon = -(a/R)$.  It will decreases as we increase the energy, for a while, and then very soon it starts to increases as energy increases at some $\varepsilon = \varepsilon_2$. Thus system will enter another phase with positive specific heat. The form of $t(\varepsilon)$ is shown in fig.\ref{t1pic}. The specific heat is positive along the portions $AB$ and $CD$ and is negative along $BC$.

\indent For $E \gg E_2 = -(Gm^2/R)$, gravity is not strong enough to keep $r < R$ and the system behaves like a gas confined by the container; we have high temperature phase with positive specific heat. As  the energy decreases to $E \leq E_2$, the effects of gravity begin to be felt. For $E_1 = -(Gm^2/R)         < E < E_2$, the system is unaffected by either the box or the short distance cutoff; this is the domain dominated entirely by gravity and we have negative specific heat. As system goes to $E \simeq E_1$, the hard core nature of the particles begins to be felt and gravity is again resisted. This give rise to a low temperature phase with positive specific heat. 

\begin{figure}
\centerline{\hbox{\psfig{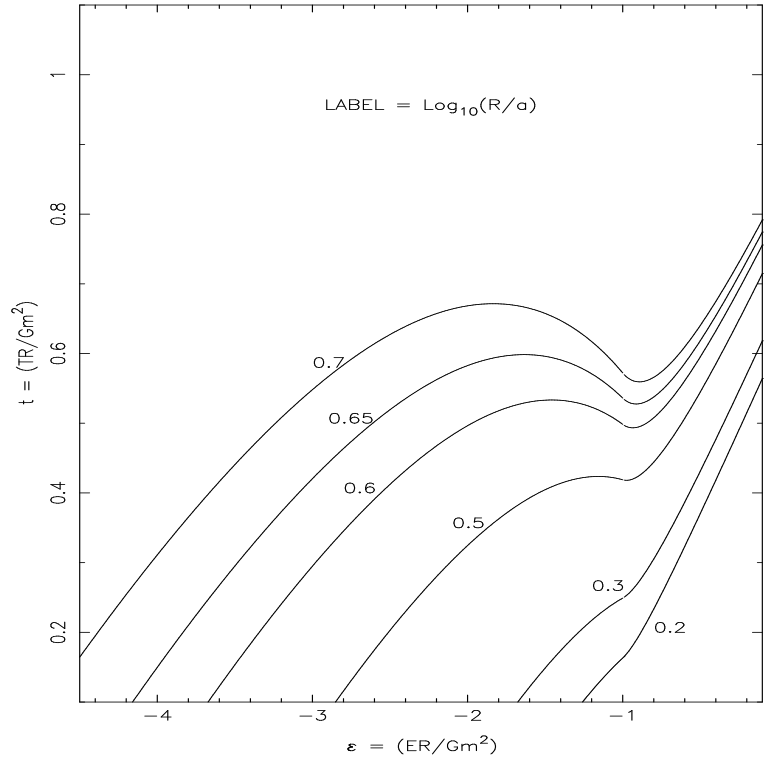}}}
\caption{ The $T(E)$ curve on the short distance cutoff $a$. For $a>(3^{1/2}+1)^{-1}R$ there is no
region of negative specific heat. As $a$ is lowered, the negative specific heat
region becomes more and more pronounced.}
\label{t2pic}
\end{figure}

\begin{figure}
\centerline{\hbox{\psfig{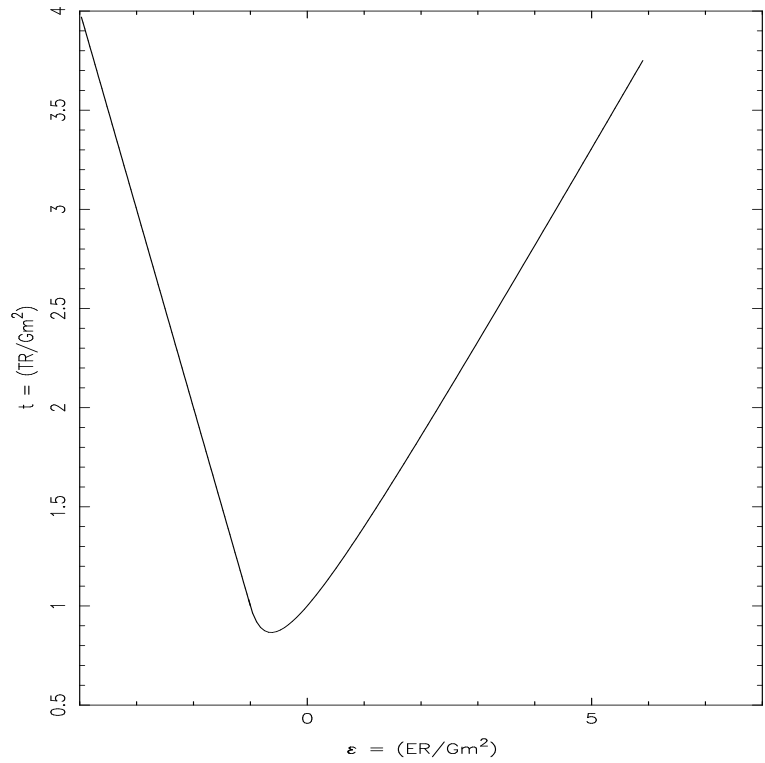}}}
\caption{ The $T(E)$ curve for $a=0$. The low temperature region with positive specific heat does not exist in the absence of the short distance cutoff. The system also exhibits a lower bound on the temperature.}
\label{t3pic}
\end{figure}

\indent It is also interesting to study (i) the effect of increasing $R$, keeping $a$ and $E$ fixed, and (ii) the effect of varying $a$, keeping $R$ fixed. In former case, it is amusing to note that, if $2<R/a<(\sqrt{3}+1)$, there is no region of negative specific heat. As we increase $R$, this negative specific heat region appears. In the latter case, at first one should rescale variables using $(Gm^2/R)$. This can be easily done and fig.\ref{t2pic} shows the behaviour of the $T(E)$ curve as the lower cutoff $a$ is changed. As $a$ is lowered, the negative specific heat region becomes more and more pronounced. If $a$ is zero, we have negative specific heat for all $E < -(Gm^2/R)$ (see fig.\ref{t3pic}).
 
\subsection{Canonical description}
 It is of interest to look at our system from the point of view of the canonical distribution. To do this we have to compute the partition function
\begin{equation}
Z(\beta) = \int d^3P\ d^3p\ d^3Q\ d^3r\ exp(-\beta H), \label{z'}
\end{equation}
\noindent which after integrating over $P, p$ and $Q$ and omitting an overall constant, which is unimportant, in dimensionless form becomes 

\begin{figure}
\centerline{\hbox{\psfig{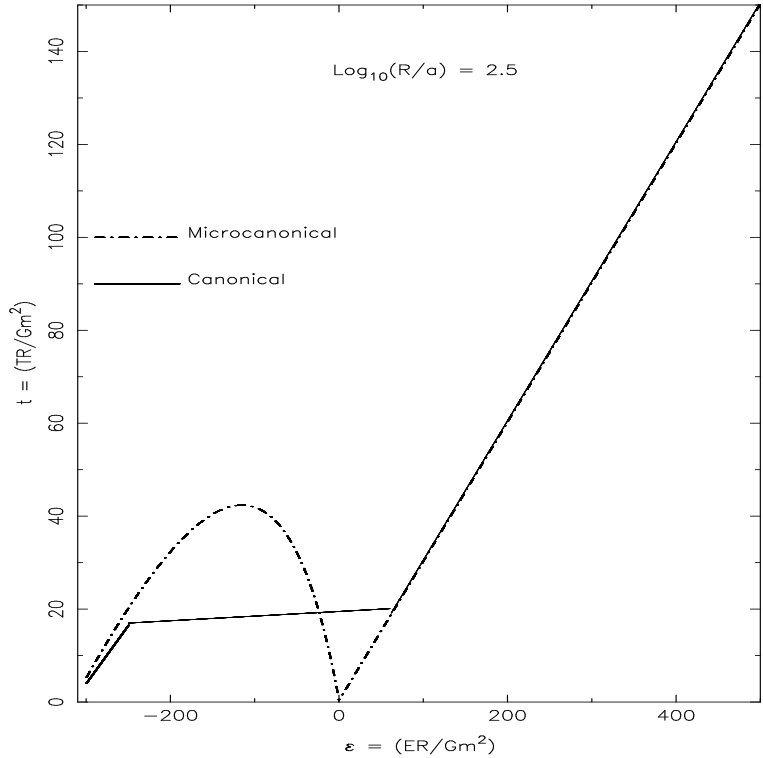}}}
\caption{ Comparison of the $T(E)$ relations for the canonical and microcanonical distributions. The negative temperature region of the microcanonical distribution is replaced by a phase transition in the canonical distribution. The microcanonical temperature is replaced by a factor $2/3$ for convenience of comparison.}
\label{t1-4pic}
\end{figure}

\begin{equation}
Z(t) = t^3(R/a)^3\int_1^{R/a} dx \ x^2 \ exp(1/xt), \label{zt}
\end{equation}
\noindent where $t$ is the dimensionless temperature defined before. One can show that $Z(t)$ can be well approximated by the expression
\begin{equation}
Z \simeq \left\{ \begin{array}{ll}
 \left (R/a \right )^3 t^4 (1-2t)^{-1} \ exp\left (1/t \right ),  & \mbox{for $t < t_c$},\vspace{5 mm}\nonumber \\
\frac{1}{3}t^3 \left (R/a \right )^6 \left (1+3a/2Rt \right ),   &  \mbox{for $t > t_c$.  \label{zappox}}               
\end{array}
\right. 
\end{equation}
\noindent where $t_c = \left [3\ln (R/a) \right ]^{-1}$, is the critical temperature at which the transition occurs. Given $Z(\beta)$ one can compute $E(\beta)$ by the relation $E(\beta) = -(\partial \ln Z/{\partial \beta})$. This relation can converted to give $T(E)$, which can be compared with the $T(E)$ obtained earlier using the microcanonical distribution. From (\ref{zappox}) we get
\begin{equation}
\varepsilon \simeq \left\{ \begin{array}{ll}
aE/Gm^2 = 4t-1, & \mbox{for $t < t_c$},\vspace{5 mm} \label{et} \\
3t-3a/2R ,     &  \mbox{for $t > t_c$.  \label{et'}}               
\end{array}
\right. 
\end{equation}
\noindent Near $t\approx t_c$ is a rapid variation of the energy and we cannot use either asymptotic form. The system undergoes a phase transition at $t = t_c$ absorbing a large amount of energy,
\begin{equation}
\Delta \varepsilon \approx 1-\frac{1}{3\ln (R/a)}.   \label{delta}
\end{equation}
\noindent The specific heat is, of course, positive throughout the range. This is to be expected because the canonical description cannot lead to negative specific heats. 

\indent The exact T(E) curves obtained from the canonical and microcanonical distributions are shown in fig.\ref{t1-4pic}. The descriptions match very well in the regions of positive specific heat. The negative specific heat region of the microcanonical distribution is replaced by a phase transition in the canonical description.

\indent The physics of canonical distribution is best understood by studying $E$ as a function of $T$. As we increase the temperature from zero to, the energy increases from the ground state value $-(Gm^2/a)$ in accordance with first equation of (\ref{et}). As the temperature approaches $T_c$ and cross it, a phase transition occurs in the system and the energy increases rapidly. The latent heat in the system is large enough to push the system into the high temperature phase. At still higher temperatures, the energy increases steadily with the temperature in accordance with second equation of (\ref{et'}).

\indent We can now compare the canonical and microcanonical descriptions for our system. At both very low and very high temperatures, the descriptions match. The crucial difference occurs at the intermediate energies and temperatures. The microcanonical description predicts a negative specific heat and a reasonably slow variation of energy with temperature. The canonical description , on the other hand, predicts a phase transition with a rapid variation of energy with temperature. Such  phase transitions are accompanied by large fluctuations in the energy, which is the main reason for disagreement between the two descriptions.

\begin{figure}
\centerline{\hbox{\psfig{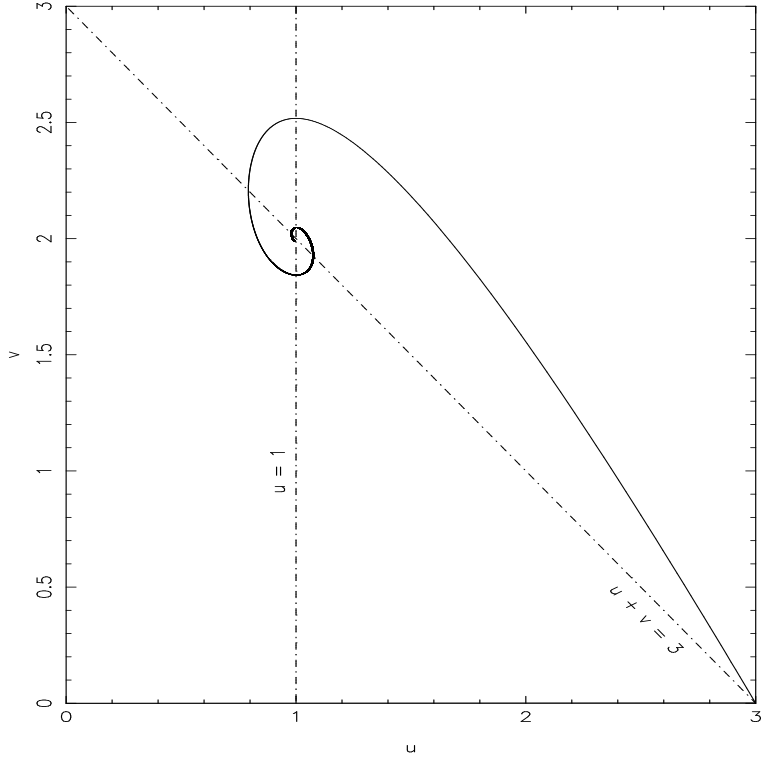}}}
\caption{ The $u-v$ curve fore the isothermal sphere without any short distance cutoff.}
\label{uvpic}
\end{figure}
In order to confirm all the conjectures made earlier based on the toy model we
shall briefly explain the equilibrium of a gravitating system in the mean field
limit. In the absence of any short distance cutoff to gravitational interaction,   the mean field solution is given by an isothermal sphere. In this case we may
write,
\begin{equation} 
\nabla^2 \phi = 4 \pi G \rho_c e^{-\beta[\phi(x)-\phi_c]}, \label{del}
\end{equation}
where $\rho_c = \rho(0)$.\\
One may introduce the following dimensionless variables in order to write the
isothermal equation,
\begin{equation}
x \equiv r/L_0,\hspace{0.5cm} n \equiv \rho /\rho_c, \hspace{0.5cm} m \equiv M(r)/M_0, \hspace{0.5cm} y \equiv \beta [\phi -\phi_c]. \label{dimless2}
\end{equation}
where $L_0$ and $M_0$ are some characteristic length and mass, respectively.
 
\noindent Then in terms of $y(x)$ the isothermal equation becomes
\begin{equation}
\frac{1}{x^2}\frac{d}{dx}\left (x^2 \frac{dy}{dx} \right ) = e^{-y}, \label{3iso}
\end{equation}
\noindent with the boundary condition $y(0) = y'(0) = 0$. By defining the new variables as
\begin{equation}
v\equiv m/x , \hspace{1cm} u\equiv nx^3/m=nx^2/v, \label{3uv}
\end{equation}
\noindent one can express the isothermal equation as the following:
\begin{equation}
\frac{u}{v}\frac{dv}{du}=-\frac{u-1}{u+v-3}, \label{3uviso}
\end{equation}
\noindent with the boundary conditions $v = 0$ at $u = 3$ and $dv/du = -(5/3)$ at $(u, v)=(3, 0)$.\\
\indent The solution curve starts at $(3, 0)$ and spirals indefinitely around the point $(1, 2)$ as $x$ tends to infinity (see fig.\ref{uvpic}). All isothermal spheres must necessarily lie on the this curve .

\indent The $u-v$ curve also implies a bound on the temperature of the system [Lynden-Bell and Wood \shortcite{lynwood}]. It is clear that for any isothermal sphere $v$ is bounded from above , i.e, $v < v_{max}$, where $v_{max}\approx 2.5$. Since

\begin{figure}
\centerline{\hbox{\psfig{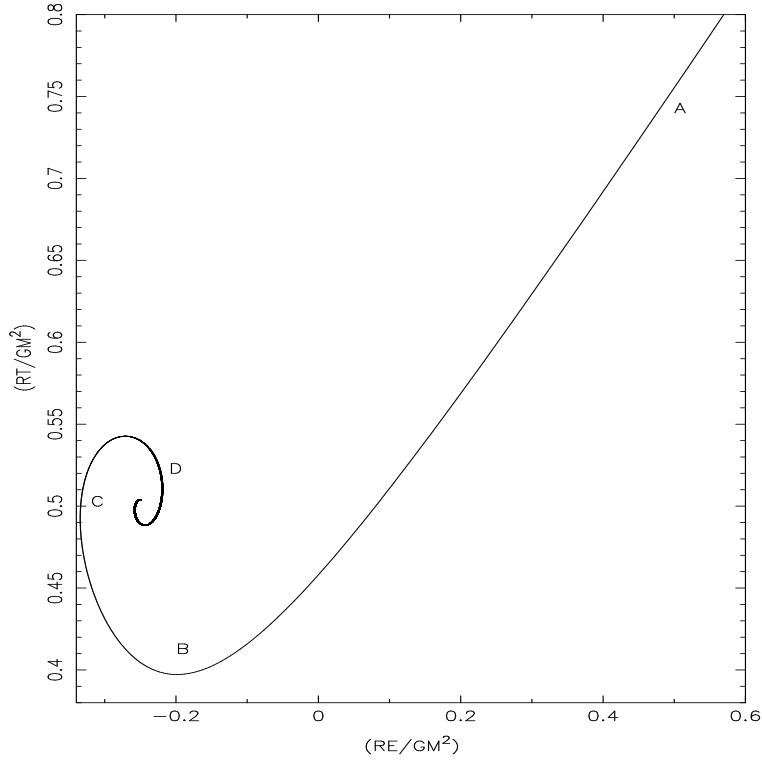}}}
\caption{ The $T(E)$ curve for the isothermal sphere without any short distance cutoff. The specific heat is positive along $AB$ and negative along $BC$.}
\label{t5pic}
\end{figure}

\begin{equation}
v = m/x = (M/M_0)(R/L_0) = (GM/R)\beta, \label{3vmax}
\end{equation}
\noindent we immediately get
\begin{equation}
T > T_{min}, \hspace{1cm} T_{min} \simeq 0.4GM^2/R. \label{3tmin}
\end{equation}
\noindent The $T(E)$ curve for the isothermal sphere can be determined by combining the relations
\begin{equation}
T = (GM^2/R)v^{-1}, \hspace{1cm} E = (GM^2/R)v^{-1}(u-\frac{3}{2}). \label{3ET}
\end{equation}
\noindent The form of the curve is shown in fig.\ref{t5pic}. The specific heat is positive along the $AB$ (which represents a high temperature phase) and is negative along $BC$. The branch $CD$ is unstable and is not physically realisable. Since we have not introduced any short distance cutoff in the system we do not have a low temperature phase with positive specific heat. If we do that, a short distance cutoff makes the curve turn to the left on the $CD$ branch and produces a second region of positive specific heat. Figure \ref{t5pic} is analogous to fig.\ref{t3pic}, which describes the $T(E)$ curve for a binary model {\it in the absence} of a short distance cutoff. The modification of fig.\ref{t5pic} in the presence of a short distance cutoff will be very similar to the modification of fig.\ref{t3pic} into fig.\ref{t2pic} (For example see fig.4.11 in Padmanabhan, 1990). 

\section{The case of 2$D$ binary}
We shall now derive corresponding results for the 2$D$ confined binary system. We will see that the thermodynamic functions are all analytically calculable for this case. It is also interesting to contrast the thermodynamical behaviour of the 3$D$ confined binary system with the different, but eventful, 2$D$ one.

\subsection{Microcanonical approach}
Let us consider two particles interacting via 2$D$ gravitational force and moving in a 2$D$ region of radius $R$. The potential in 2$D$ gravity satisfies the Poisson equation:
\begin{equation}
\nabla^2\phi = 2\pi G\rho. \label{poisseq}
\end{equation}
\noindent For point particle $\phi$ will be logarithmic; the potential energy of interaction between two such particles will be,
\begin{equation}
U({\bf x_1}, {\bf x_2}) = Gm^2\ln\frac{|{\bf x_2}-{\bf x_1}|}{R}. \label{potenergy}
\end{equation}

\begin{figure}
\centerline{\hbox{\psfig{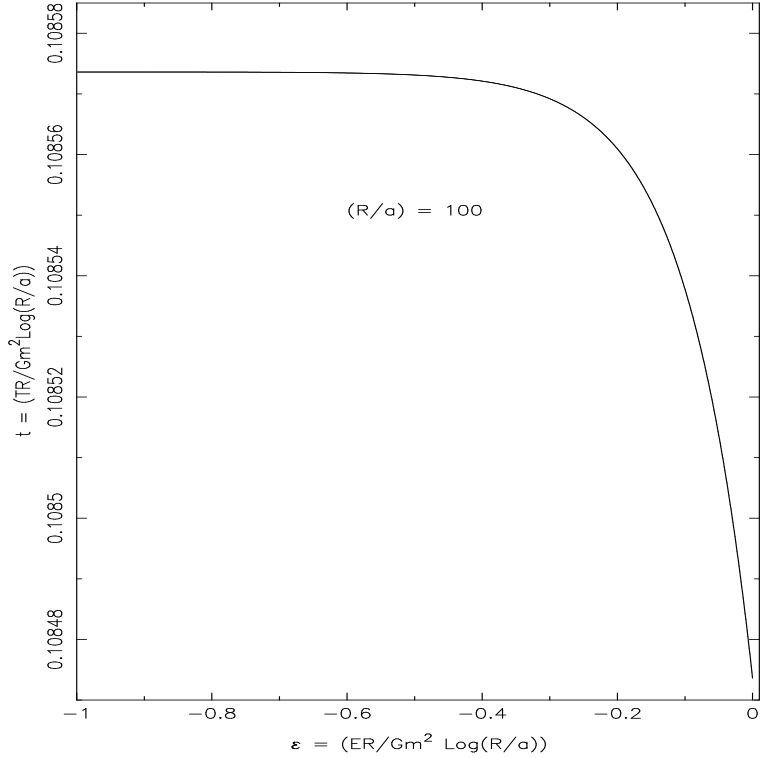}}}
\caption{ The $T(E)$ curve for the low energies region for 2$D$. In spite of existence of short distance cutoff, there is a region which specific heat is negative, although very small.}
\label{t6pic}
\end{figure}

\begin{figure}
\centerline{\hbox{\psfig{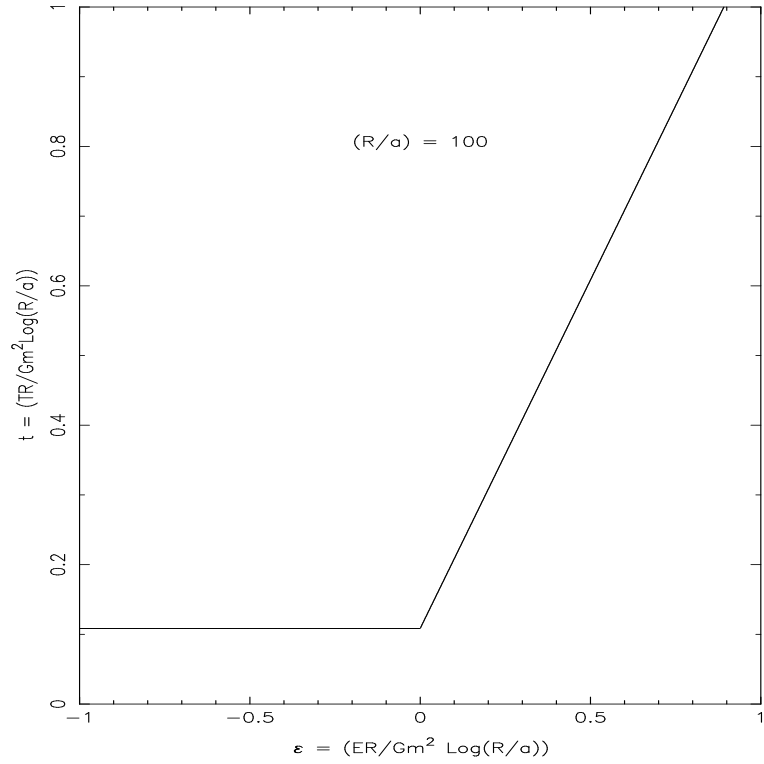}}}
\caption{ The $T(E)$ curve. Note that the negative specific heat region is
not visible due to its smallness compare to high energy region.}
\label{t7pic}
\end{figure}

We shall first study the microcanonical distribution corresponding to our system. The phase volume $g(E)$, which is the volume of the constant energy surface, in the phase space:
\begin{equation}
g(E)  = \int \prod_{i = 1}^2 d^2x_id^2p_i\delta(E-H), \label{phasespace}
\end{equation}
\noindent where, $H$, is the Hamiltonian for a 2$D$ gravitating system of 2 particles with logarithmic potential, defined as,
\begin{equation}
H = \frac{{\bf P^2}}{2M}+\frac{{\bf p^2}}{2\mu}+Gm^2\ln\left(\frac{r}{R}\right), \label{ham2}
\end{equation}
\noindent where, as in the 3$D$ case, $({\bf Q}, {\bf P})$ are the coordinates and momenta of the centre of mass, $({\bf r}, {\bf p)}$ are the relative coordinates and momenta, $M = 2m$, is the total mass, $\mu = (m/2)$ is the reduced mass and m is the mass of the individual particles. we shall further restrict the range of the coordinate $r$ to the interval $(a, R)$. Where, $a$, is the short distance cutoff equivalent to assuming that particles are hard spheres of radius $(a/2)$, while, $R$, is the large distance cutoff equal to the confining radius of the system. Therefore, the phase volume $g(E)$ becomes,
\begin{equation}
g(E) = AR^2\int_a^{r_{max}}r\left[E-Gm^2\ln\left(\frac{r}{R}\right)\right] dr. \label{phasevol}
\end{equation}
\noindent With the range of integration in (\ref{phasevol}), limited to the region in which the expression in parentheses is positive, i.e., $  [E-Gm^2\ln{(r/R)}] > 0$. Therefore, 
\begin{equation}
{r_{max}} = \left\{ \begin{array}{ll}
R \ \exp{(E/Gm^2)},  & \mbox{$-Gm^2\ln(R/a) < E < 0$},\vspace{5 mm}\nonumber \\
R,   &  \mbox{$0 < E < \infty$.  \label{rmax}}               
\end{array}
\right. 
\end{equation}
\noindent $A$ is some constant which is irrelevant to our discussion.

\indent The integration, then, will yield the following result:

\begin{figure}
\centerline{\hbox{\psfig{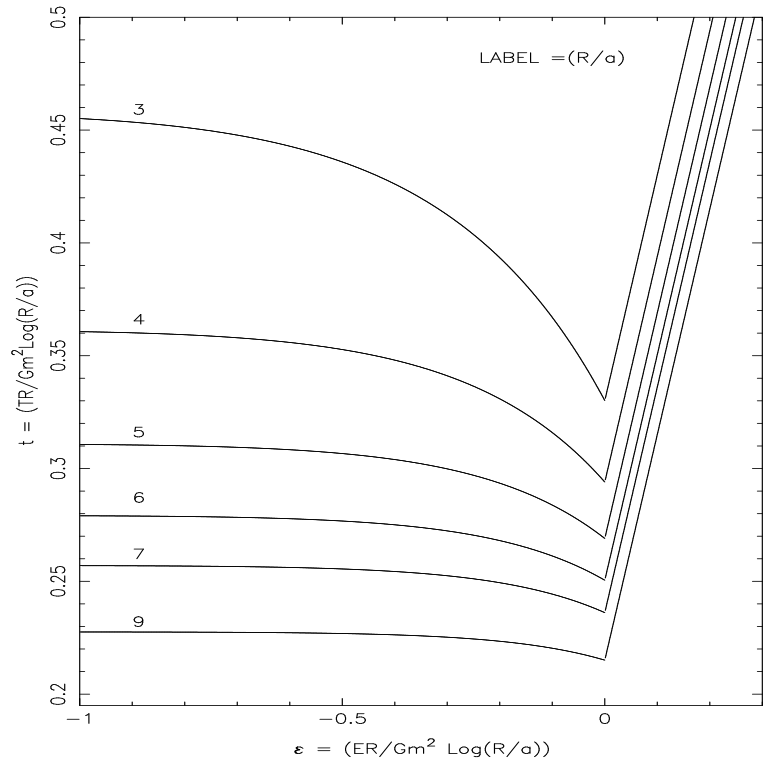}}}
\caption{ The $T(E)$ curve for different short distance cutoff $a$. As one decreases $R$, the negative specific heat will become more and more pronounce.}
\label{t8pic}
\end{figure}

\begin{figure}
\centerline{\hbox{\psfig{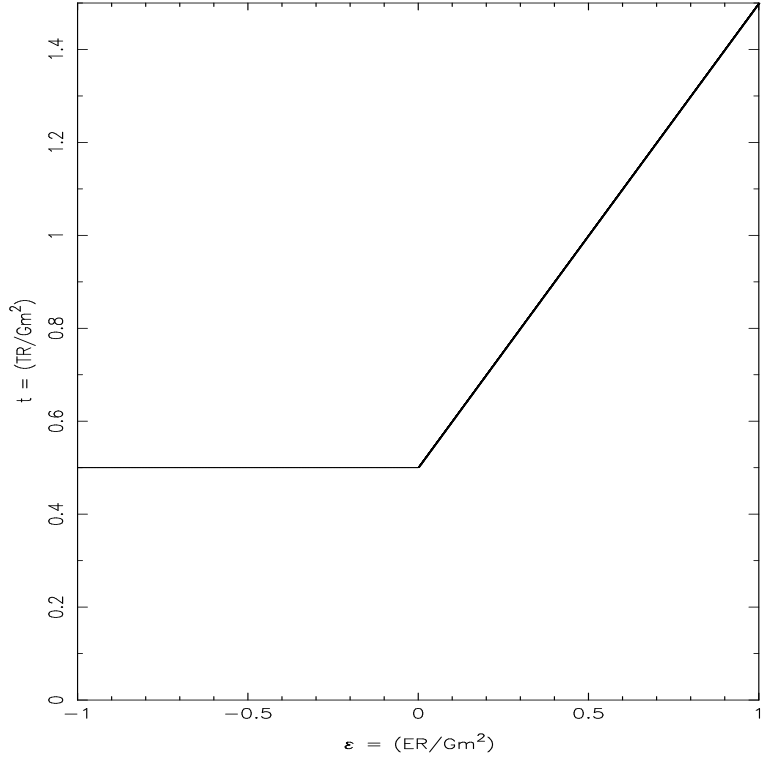}}}
\caption{ The $T(E)$ curve for the special case, $a = 0$. The system does not exhibit any negative specific heat. Compare this figure with the fig.3.}
\label{t10pic}
\end{figure}

\begin{equation}
{\frac{g(E)}{Gm^2}} = \left\{ \begin{array}{ll}
\frac{1}{4}AR^2\{R^2e^{2E/Gm^2}-a^2(2E/Gm^2+2\ln(R/a)+1)\},  & \mbox{$-Gm^2\ln(R/a)<E<0$},\vspace{5 mm}\nonumber \\
\frac{1}{4}AR^2\{R^2(2E/Gm^2+1)-a^2(2E/Gm^2+2\ln(R/a)+1)\}.   &  \mbox{$0<E<\infty$.  \label{huge2}}               
\end{array}
\right. 
\end{equation}
\noindent It is obvious that $g(E)$ is continuous and smooth at $E = 0$. We shall study the thermodynamics of the system, which can be now studied using $g(E)$, in the two regimes given above.

The entropy of the system, $S(E) = \ln{g(E)}$ in the case of very low energies, i.e., $Gm^2 \ln (a/R) < E < 0$ will be,
\begin{equation}
S(E)  = \ln\left[\frac{1}{4}AGm^2R^2\right]     +\ln\left[R^2e^{2E/Gm^2}-a^2\left(\frac{2E}{Gm^2}+2\ln\left(\frac{R}{a}\right)+1\right)\right],\label{ent}
\end{equation}
\noindent and the temperature of the system, $T(E)^{-1} = \partial S(E)/\partial E$ with the help of (\ref{ent}) is,
\begin{equation}
T(E) = Gm^2\left (\frac{1}{2}-\frac{[E/Gm^2+\ln(R/a)]}
           {(R/a)^2e^{2E/Gm^2}-1}\right ), \label{temper}
\end{equation}
\noindent or in dimensionless form as,
\begin{equation}
t(\varepsilon) = \frac{1}{2\ln(R/a)}+\frac{(\varepsilon +1)}
 	  	{1-(R/a)^{2(1-\varepsilon)}}, \label{t}
\end{equation}
\noindent where we have defined $t(\varepsilon) = T(E)/{Gm^2\ln(R/a)}$, and $\varepsilon = E/{Gm^2\ln(R/a)}$. At the lowest energy admissible for our system, which corresponds to $\varepsilon = -1$, the temperature is
$t(\varepsilon)=[2\ln(R/a)]^{-1}$. It is clear that in (\ref{t}) for $\varepsilon \simeq -1$ the first term dominates. So as we increase the energy of the system, the temperature {\it decreases}. This behaviour continues up to $\varepsilon = 0$ at which the point $t(\varepsilon)$ curve reaches to it's minimum, $t(\varepsilon) = 1/\{2\ln(R/a)\}-a^2/(R^2-a^2)$. Therefore, we obtain a negative specific heat region for $-1 < \varepsilon < 0$ (see fig.\ref{t6pic}).

For $E \geq 0$ we should use the second expression in (\ref{huge2}) for $g(E)$. In this case the proper expression for $S(E)$ would be,
\begin{equation}
 S(E) =  \ln\left[\frac{1}{4}AGm^2R^2\right]
            +\ln\left[R^2\left(\frac{2E}{Gm^2}+1\right)-a^2\left(\frac{2E}{Gm^2}+2\ln\left(\frac{R}{a}\right)+1\right)\right], \label{entro}
\end{equation}
\noindent and thus we get

\begin{equation}
t(\varepsilon)=\varepsilon+\frac{1}{2\ln(R/a)}+\frac{1}
 	  	  		 {1-(R/a)^2}. \label{t'}
\end{equation}

\noindent This function, clearly, matches with (\ref{t}) at $\varepsilon = 0$. As we increase the energy , the temperature continues to increase. Thus in high temperature phase we enter to the positive specific heat region. The form of $t(\varepsilon)$ is shown in fig.\ref{t7pic} (Due to smallness of negative specific heat region compared to high energy limit, the negative specific heat region is not visible in the figure). However, as we decrease the long distance cutoff $R$, the negative specific heat region becomes more and more pronounce, see fig.\ref{t8pic}. \\ 
\indent If we set $a = 0$, we easily get,
\begin{equation}
\frac{T(E)}{Gm^2} = \left\{ \begin{array}{ll}
1/2, & \mbox{$-\infty<E<0$}, \vspace{5 mm}\nonumber \\
1/2+E/Gm^2, & \mbox{$0<E<\infty$}. \label{RT}
\end{array}
\right. 
\end{equation}

It is obvious that in the absence of the short distance cutoff, there will
be no negative specific heat region(fig.\ref{t10pic}). Comparing this result with it's counterpart in 3$D$ case for $a = 0$ shows that negative specific heat replaced
by constant temperatures. In fact, the effect of short distance cutoff in 2$D$ case is just destablising the effect of gravitational potential energy, unlike its effect
in 3$D$ potential. By putting a short distance cutoff, we actually distort
the phase space. In other words, having a short distance cutoff in the
system is equivalent to removing some part of phase space. This distortion
in phase space may cause many unexpected consequences. In 2$D$, for instance,
removing some part of phase space causes an unexpected region of negative
specific heat, whereas for 3$D$, we get region of positive specific heat.
Comparison of 2$D$ and 3$D$ with short distance cut off, though, suggests
the $E-T$ graph in 2$D$ in many aspects is similar to its 3$D$ counterpart,
except that in 2$D$ there is no initial positive specific heat region. In
other words, the $E-T$ graph in 2$D$ is almost same as the $E-T$ graph
in 3$D$, but as if the graph has been shifted to the right: the starting
point of the 2$D$ graph  almost coincides with the maximum of 3$D$ one.

\subsection{Canonical approach}
Let us now consider the partition function $Z(\beta)$ of the system, which is given by the integral,
\begin{equation}
Z(\beta)=\int_{-\infty}^{+\infty} dE \ g(E) e^{-\beta E} = \int_{-\infty}^0 dE \ g(E) e^{-\beta E} + \int_0^{+\infty} dE \ g(E) e^{-\beta E} \equiv Z_1+Z_2\label{par}
\end{equation}
\noindent The range of integration is from $(-\infty)$ to $(+\infty)$, since
negative values of $E$ are allowed. Thus in our case, that interval will break to two separate areas. We have for $Z_1$:
\begin{equation}
Z_1(\beta) = \frac{1}{4}AR^4Gm^2\int_{-\infty}^0 dE  \  exp\left ([2/Gm^2-\beta]E\right )
         \label{parti}
\end{equation}
\noindent The above integrand will diverge in the lower limit, unless,
\begin{equation}
\beta < \frac{2}{Gm^2}. \label{beta}
\end{equation}
\noindent Then we obtain,
\begin{equation}
Z_1(\beta) = \frac{1}{4}AR^4\frac{(Gm^2)^2}{2-Gm^2\beta}. \label{partii}
\end{equation}
\noindent  Since $Z_1(\beta)$ diverges at $Gm^2\beta_c = 2$, the system can exist only  at $\beta < \beta_c$. As for $Z_2$,
\begin{eqnarray}
Z_2(\beta) & = & \frac{1}{4}AR^4\int_0^{\infty} dE (Gm^2+2E)e^{-\beta E}
                \nonumber \\
           & = & \frac{1}{2}A\left ( \frac{R^2}{\beta}\right )^2 \left ( 1+\frac{\beta Gm^2}{2} \right ), 
                 \label{z}
\end{eqnarray}
\noindent which is the $Z(\beta)$ as one can obtain from saddle-point approximation \cite{paddy}. Thus, we see that the saddle-point approximation is accurate at high temperatures, i.e. for $Gm^2\beta \ll 1$, and the mean field
approximation must break down at low temperatures.
If now we introduce the short distance cutoff, $a$, the corresponding
$Z_1$ and $Z_2$ will be modified as

\begin{equation}
Z_1(\beta)=\frac{1}{4}AR^4\frac{(Gm^2)^2}{2-Gm^2\beta}
-\frac{1}{2} A \left(\frac{Ra}{\beta}\right)^2\left[
1+\frac{1}{2}Gm^2\left[2\ln{(R/a)}+1\right]\beta\right], \label{partii1}
\end{equation}
and
\begin{eqnarray}
Z_2(\beta) & = & \frac{1}{4}AR^4\int_0^{\infty} dE (Gm^2+2E)e^{-\beta E}
                \nonumber \\
         & = & \frac{1}{2}A\left ( \frac{R^2}{\beta}\right )^2 \left ( 1+\frac{\beta Gm^2}{2} \right )-\frac{1}{2} A
\left(\frac{Ra}{\beta}\right)^2\left[
1+\frac{1}{2}Gm^2\left[2\ln{(R/a)}+1\right]\beta\right]. 
                 \label{z2}
\end{eqnarray}
\noindent Given $Z(T)$, one can compute the mean energy of the system, which is given
by
\begin{equation}
E(T) = T^2(\partial \ln{Z}/\partial T),
\end{equation}
or in dimensionless form as
\begin{equation}
\varepsilon = t^2 (\partial \ln{Z}/\partial t),
\end{equation}
where for high energies, regardless of whether there is a short distance
cutoff, the dimensionless energy scales as
\begin{equation}
\varepsilon_2 \simeq \frac{2t^2}{2t+1}\sim t.
\end{equation}
\noindent However at $t\approx t_c = 1/[2\ln(R/a)]$, energy diverges.
Thus, again at $Gm^2\beta_c=2$, the partition function will blows up
even in the presence of short distance cutoff $a$ and in this
case the system cannot exist at $\beta > \beta_c$, as well. In
other words, the canonical description of the system exists only at sufficiently high temperatures \cite{kalyn}. Note that there is no phase transition for this
two dimensional system either with or without short distance cutoff. As the energy of the system is lowered, the temperature continuously decreases
and asymptotically approaches $T_c = (1/\beta_c)$.

Comparison of the canonical and microcanonical descriptions of our model
shows that at very high temperatures, the descriptions match. The main
difference occurs at low temperatures and energies. The microcanonical description predicts either {\it infinite} or {\it negative} specific heat, depend on
whether there is a short distance cutoff, at low energies, whereas
in canonical approach, there is no physical state below a critical
temperature. Therefore, for low energies the two approaches disagree.

\begin{figure}
\centerline{\hbox{\psfig{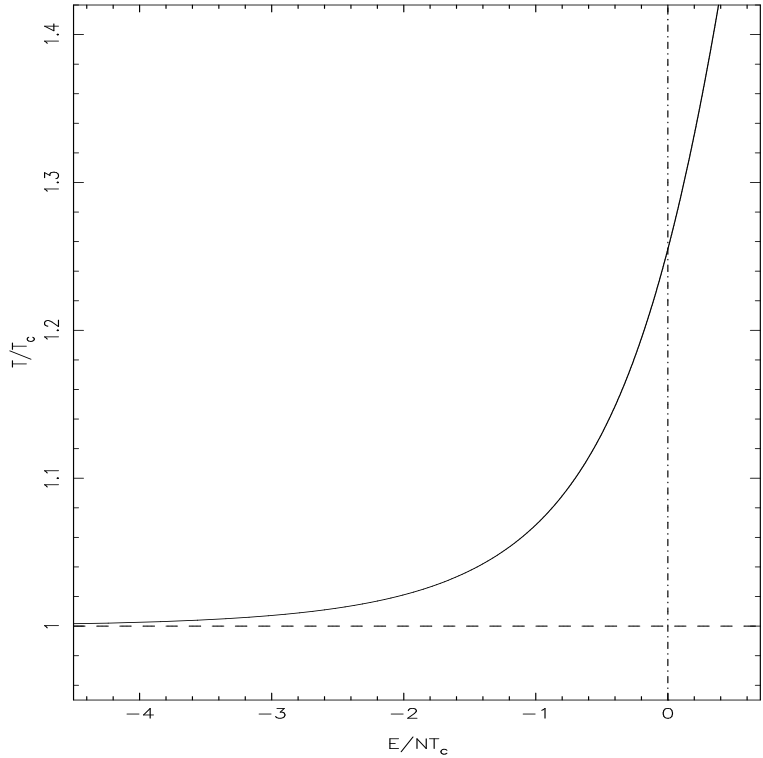}}}
\caption{ The $T/T_c$ as a function of the total energy of two dimensional gravitating system. There is no region of negative specific heat but the system exhibits a lower bound on the temperature.}
\label{t11pic}
\end{figure}

As we compared the result of 3$D$ binary with isothermal sphere, we shall
outline the result of studying isothermal cylinders in order to confirm our
earlier claim based on 2$D$ binary.

An isothermal self-gravitating cylinder 
is described by Poisson's equation, (\ref{poisseq}), in 2$D$,
\begin{equation}
\frac{1}{r}\frac{d}{dr}\left (r \frac{d\varphi}{dr}\right ) = 2 \pi G \rho(r), \label{2dpoisseq}
\end{equation}
\noindent with $\rho(r)$ given by
\begin{equation}
\rho (r) = A exp[-\beta \varphi(r)],
\end{equation}
\noindent where the constants $A$, and $\beta$ are to be determined in terms of total mass $M$ and energy $E$ of the system. Regular solution of (\ref{2dpoisseq}) have been given in Ostriker \shortcite{jerry}, and   Stodolkiewicz \shortcite{stod}:
\begin{equation}
\varphi(r) = GM \ln{R} + 2 \beta^{-1} \ln\left[1-\frac{1}{4}GM\beta (1-r^2/R^2)\right]+ constant , \label{exactpot}
\end{equation}
\noindent where $R$ is the radius of the confining box. The potential on the axes, $R = 0$, is real if $(1/4)GM\beta$ is smaller than unity and becomes infinite in the limit $(1/4)GM\beta = 1$. Thus, there exist a lower bound, $T_{c}$.
Now all other physical variables like density $\rho(r)$, pressure on the confining wall $P(R) = \beta^{-1} \rho(r)$, and the energy
\begin{equation}
E = M\beta^{-1} +\frac{1}{2} \int \rho \varphi d{\bf x}, \label{2energy}
\end{equation}
\noindent can be computed from (\ref{exactpot}). We get
\begin{equation}
E = \frac{1}{4}GM^2\left[2\ln R +2 (T/T_c)+(T/T_c)^2\ln(1-T_c/T)+contstant\right], \label{exacee}
\end{equation}
\begin{equation}
PV = N(T-T_c), \label{pv}
\end{equation}
\noindent where $T_c = (1/4)NGM^2$ is the same critical temperature as that found earlier (for $N = 2$) in the canonical approach. It is clear the system can not exist for $T < T_c$, the pressure becomes negative, and the potential at the origin diverges.

\indent The $T(E)$ curve for this system is shown in fig.\ref{t11pic}. It is clear that there is no region of negative specific heat for this two dimensional system. As $E$ is lowered, the $T(E)$ continuously decreases, however, $T$ will never reach $T_c$ since for that to happen, an infinite amount of energy has to be given away. As $T$ tends to $T_c$ the pressure on the wall tends to zero, see equation (\ref{pv}), the density in the centre grows continuously as
\begin{equation}
\rho(0) = \frac{M}{V}(1-T_c/T)^{-1}, \label{rho0}
\end{equation}
\noindent and the density contrast grows as,
\begin{equation}
\rho(0)/{\rho(R)} = (1-T_c/T)^{-2}.
\end{equation}
\noindent The system, after shrinking, collapses to thin dense {\bf string}. Therefore, isothermal
cylinders in contact with heat bath whose temperature $T$ is slightly smaller than $T_c$ are unstable and giving up an unlimited amount of energy \cite{kalyn}. However, a comparison with the case of 3$D$ binaries
shows that, two-dimensional systems are more stable than three-dimensional ones.
As we obtained in eq. (\ref{3tmin}) there also exists a $T_{c}$ for isothermal spheres, in which there is no equilibrium. If an isothermal sphere has a density contrast less than 32 \cite{lynwood} and temperature slightly hotter than
$T_{c}$, while the density contrast grows, it will lose energy to heat bath
and cool down. As the density contrast keeps growing above 32, the specific
heat becomes negative. When density contrast reaches the value 709 \cite{kalyn}
and becomes very hot, the system is unstable and not physically realisable.
Then after, isothermal starts collapsing in which the centre of the system
becomes smaller and hotter whereas giving up energy to the outside parts
of the isothermal sphere.

Instabilities in isothermal spheres are mainly due to  ``{\it wiggling}'' of
potential around singular solution (see, e.g. fig.6). There is no singular
solution and thus wiggling of potential in 2$D$ (see fig.12).

\begin{figure}
\centerline{\hbox{\psfig{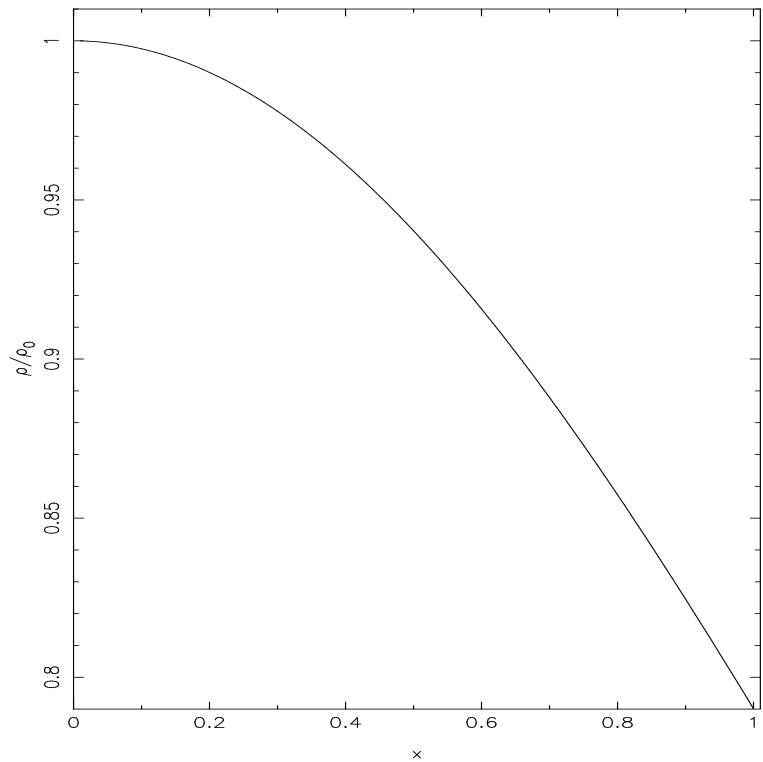}}}
\caption{ The isothermal cylinder density curve. There is no singularity
in contrast to isothermal sphere density.}
\label{t13pic}
\end{figure}

\section{conclusion}
It seems a binary system exhibits several important properties of more complicated gravitating systems in spite of the fact it has only two degrees of freedom. In particular, this system exhibits the following two features, which seems to be generic to all gravitating systems: (i) When studied using the microcanonical ensemble, the system shows evidence for two different phases: a high temperature phase, dominated by kinetic energy, and a low temperature phase dominated by the potential energy and (de)stabilised by some short distance cutoff in (2)3$D$ which is of non-gravitational origin, mainly due to distortion of phase space.  Both these phases have positive specific heat in 3$D$ whereas in 2$D$ the latter phase has negligible negative specific heat for the case of $(a/R) \ll 1$. In 3$D$ case, these two phases are connected at intermediate temperatures by a region of negative specific heat; this is precisely the range in which the kinetic and potential energies are comparable and the system is in virial equilibrium. (ii) If the system is studied using canonical ensemble, the intermediate region of negative specific heat in 3$D$ is replaced by a sharp phase transition releasing a large amount of latent heat. This suggests the following analogy: Gravitating systems in virial equilibrium are similar to normal systems (with short range forces) at the verge of phase transition. For
the case of 2$D$, however, the canonical description leads to the completely
different picture in low energies. The system does not exist below some
critical picture $T_c = (1/2)Gm^2$.

\indent On the other hand, the isothermal considerations reveal the similarity of results in this context with these simple toy model binaries. In fact, the mean field analysis confirms all the conjectures made earlier based on the binaries. 

\section*{ACKNOWLEDGMENT}
This work was done under the guidance of T.~Padmanabhan.

\end{document}